\documentclass{aa}  

\usepackage{graphicx}
\usepackage{color}
\usepackage{subfigure}
\usepackage{ulem}
\usepackage{txfonts}
\def\me{$\,{\rm M}_{\oplus}\,$}
\def\h2o{H$_2$O}
\def\sio2{SiO$_2$}
\def\gc3{g\,cm$^{-3}$}
\def\swr{Schwarzschild }
\def\ldx{Ledoux }
\definecolor{gry}{gray}{0.4}

\begin{document}

   \title{Jupiter's evolution with primordial composition gradients}

   \author{Allona Vazan
          \inst{1}
          \and
          Ravit Helled
          \inst{2}
          \and 
          Tristan Guillot
          \inst{3}
          }

   \institute{Anton Pannekoek Institute for Astronomy, University of Amsterdam\\
    Science park 904, 1098 XH, Amsterdam, the Netherlands.\\
              \email{a.vazan@uva.nl}
         \and
             Institute for Computational Science\\
Center for Theoretical Astrophysics \& Cosmology 
University of Zurich\\
Winterthurerstr. 190, CH-8057 Zurich, Switzerland.\\
             \email{rhelled@physik.uzh.ch}
             \and
             Observatoire de la Cote dAzur, Bd del'Observatoire, CS 34229, 06304 Nice Cedex 4, France. \\
             \email{tristan.guillot@oca.eu}
             }


  \abstract
{Recent formation and structure models of Jupiter suggest that the planet can have composition gradients and not be fully convective (adiabatic). 
This possibility directly affects our understanding of Jupiter's bulk composition and origin. 
In this Letter we present Jupiter's evolution with a primordial structure consisting of a relatively steep heavy-element gradient of 40\me. 
We show that for a primordial structure with composition gradients, most of the mixing occurs in the outer part of the gradient during the early evolution (several 10$^7$ years), leading to an adiabatic outer envelope (60\% of Jupiter's mass).  
We find that the composition gradient in the deep interior persists, suggesting that $\sim$40\% of Jupiter's mass can be non-adiabatic with  a higher temperature than the one derived from Jupiter's atmospheric properties.
The region that can potentially develop layered-convection in Jupiter today is estimated to be limited to $\sim$10\% of the mass.  
}

   \maketitle

\section{Introduction}

Determining the heavy-element mass (hereafter, M$_Z$) and its distribution in Jupiter plays a key role in understanding its origin \citep[e.g.,][]{helledlun14}. 
Typically, M$_Z$ in Jupiter is inferred by structure models that fit the observed properties of the planet, in particular, the gravitational moments $J_{2n}$ \citep{nettel15,miguel16,wahl17}.
Structure models are relatively complex as they try to reflect various physical processes and properties within the planet such as the planetary core and helium separation from hydrogen (helium rain). In addition, structure models must rely on state-of-the-art equations of state to derive the planetary composition and its depth dependence.  
Nevertheless, most interior models are derived under the simplifying assumption that Jupiter is mostly adiabatic \citep[e.g.,][]{guillot99,guilsteven04,fortney11b,nettel15}. 
Some models assume that the heavy elements are homogeneously mixed within the planetary envelope, while others allow the heavy element mass fraction (hereafter, $Z$) to change between the two regions of the envelopes created by helium-rain \citep{forthub04,nettel15}.
Although the idea that Jupiter could have composition gradients and be non-adiabatic has been proposed decades ago \citep{stevenson82a,stevenson85}, only recently this possibility has received more attention in both formation models \citep{lozovs17,helledsteven17} and structure modes \citep{lecontechab12,vazan16}. 
\par

Recently, we showed that shallow composition gradients as suggested by \cite{lecontechab12} which allow the planet to consist of more metals are not stable against convection and mixing \citep[][hereafter VHPK16]{vazan16}. 
In this case the heavy-elements mix with the hydrogen-helium envelope within a few 10$^7$ years, leading to a fully-mixed Jupiter. As a result, we suggested that when considering a non-adiabatic structure for giant planets, one must follow the long-term evolution of the planet and confirm that the current-state structure is consistent with the planetary cooling/contraction history.   
Below we present the evolution of Jupiter with primordial composition gradients (diluted core) that lead to a structure that is consistent with observations. 

\section{Evolution Model}

For calculating M$_Z$ and the internal structure of Jupiter today, we search for evolution models that lead to a current-state Jupiter model that fits the observed parameters. 
It should be noted that there are many possible initial configurations and evolutionary paths that fail, and can therefore be excluded. Here we present one possible Jupiter model that is compatible with the observational constraints. This model is used to demonstrate the efficiency of mixing and the expected evolution of giant planets with primordial composition gradients.  
The nature of evolution models is different from those of structure models, and they cannot be as accurate, therefore for the observational constraints we use Jupiter's mass, radius, effective temperature and $J_2$. Recently, $J_2$ has been determined by Juno (in units of 10$^6$) to be 14,696.514$\pm$ 0.272 \citep{folkner17}, but here we allow $J_2$ to fit within 1\%. For investigating the long-term evolution of Jupiter such an accuracy is sufficient \citep{nettel17}. Our current-state Jupiter model has a mean radius of 69,911 km and an effective temperature T$_{eff}$ = 124.6 K, as constrained by observations.  
\par  

We use the SCVH equation of state (EOS) for hydrogen and helium \citep{scvh} and our own EOS for water calculated using the QEOS method \citep{vazan13}. 
For the calculation of the heavy-element mass we assume that the heavy elements are all in \h2o. While this is a simplification, the derived $M_Z $ is not very sensitive to the assumed heavy element, although the efficiency of mixing is expected to be smaller for heavier materials (VHPK16). 
The derived $M_Z $, however, depends on the assumed hydrogen and helium EOS \citep{saum+guil04,miguel16}.

The model has a gray atmosphere with an albedo of A=0.343 \citep{guillgaut14} and includes stellar irradiation \citep[with T$_{irr}$=110K, see Appendix A3 in][for details]{vazan15}.
The opacity is set by the harmonic mean of the conductive and radiative opacities.   
The radiative opacity is the analytical fit of \cite{valencia13} to the opacity tables of \cite{freedman08}, and the conductive opacity is taken from from \cite{potekhin99}. 
More details on the evolution model are given in the appendix.

\section{Results}
Formation models including planetesimal dissolution generally lead to a primordial Jupiter that contains a relatively steep heavy-element gradient in which only the innermost region is of pure Z \citep{lozovs17,helledsteven17}.
For such a primordial structure, which also leads to a current-state structure that is consistent with Jupiter observables, 
we find the total $M_Z$ to be 40\me, when considering SCVH EOS for hydrogen and helium.
In our primordial structure, the pure-heavy-element region has a mass of only 0.3\me, with almost 40 \me within the gradient. 
At present day, due to convective mixing at early times, the gradient consists of only 24\me with $\sim$ 16\me of heavy elements being homogeneously mixed in the outer part of the planet. This current state structure is consistent with Jupiter observables.

The primordial temperature profile has a critical role in determining the planetary evolution and current-state structure in the presence of composition gradients. 
Primordial central temperatures of several 10,000\,K lead to very efficient convective-mixing at early times, and thus to a homogeneously-mixed planet. On the other hand,  
for primordial central temperatures lower than 20,000\,K convective-mixing is  negligible, and the structure is expected to be unchanged during the evolution (i.e., no mass redistribution). 
Since most giant formation models predict primordial central temperatures of 3-7$\times10^4$ \citep{mordasini12}, convective-mixing is likely to be an important mechanism in young giant planets. 
It is important to note that most of the mixing occurs during the early evolution (up to the first few 10$^7$ years). 
The outer $\sim$ 60\% (by mass) of Jupiter becomes homogeneously mixed with Z$\sim$8\%. 
At some point the luminosity in the deep interior decreases and heat is transferred by conduction. Then small stairs are created, but the further mixing of heavy elements in negligible.

\subsection{The Evolution of the Internal Structure}
   \begin{figure}
\centerline{\includegraphics[angle=0, width=9.6cm]{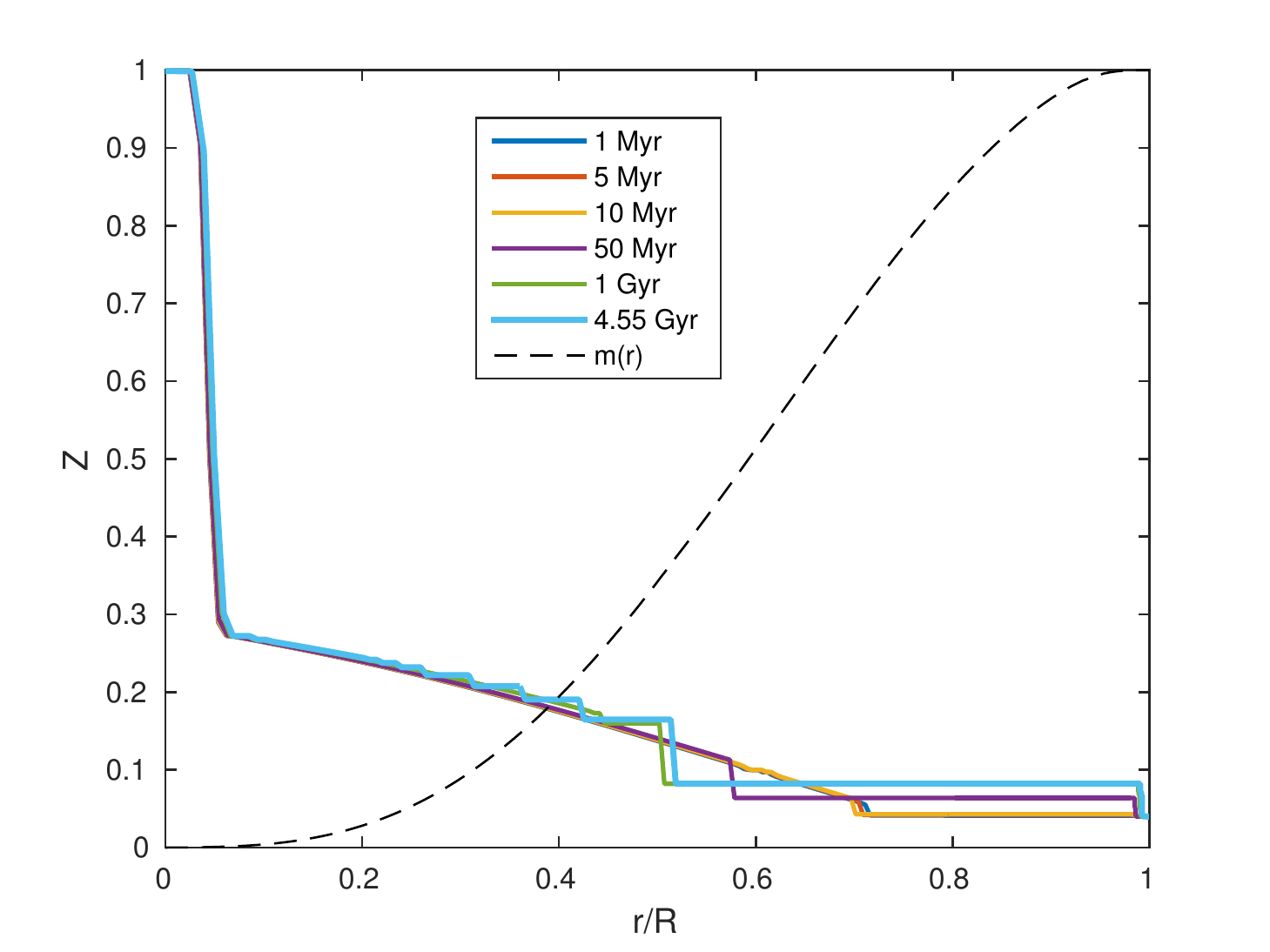}}
\vspace{-0.8ex}
\centerline{\includegraphics[angle=0, width=9.6cm]{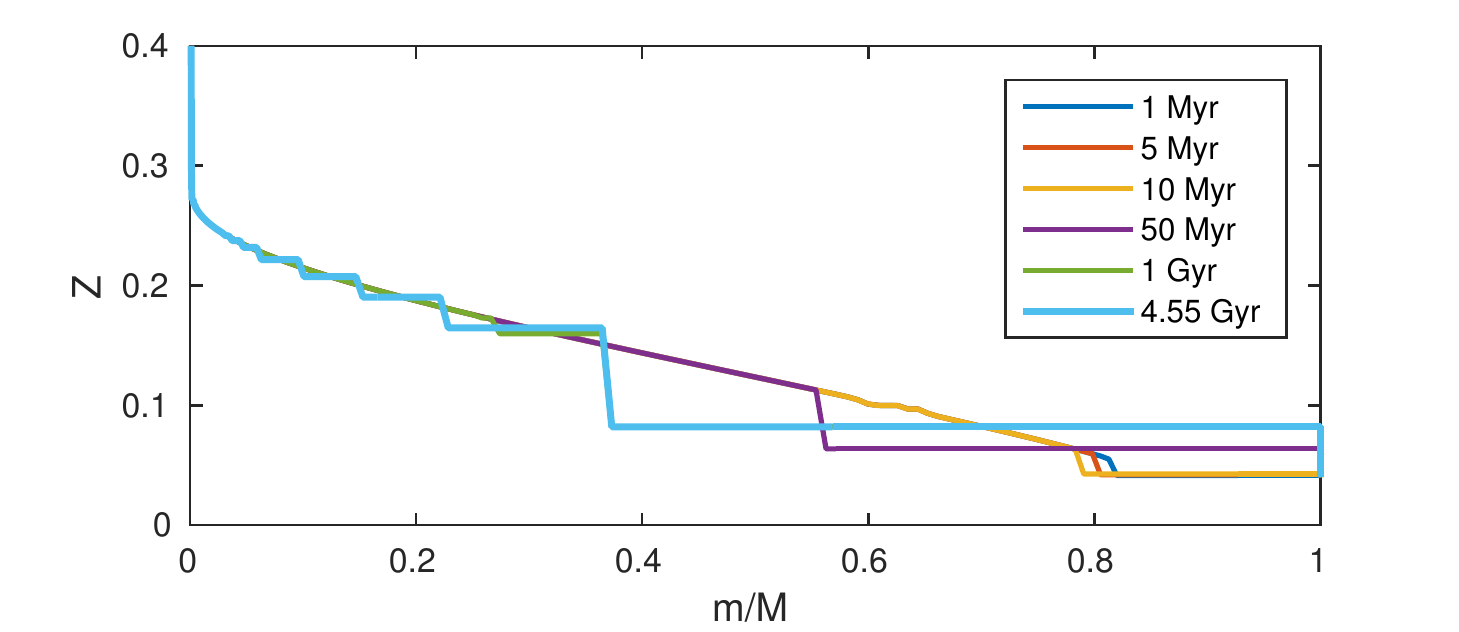}}
\vspace{-1.0ex}
\caption{The heavy element mass fraction Z vs. normalized radius (top) and mass (bottom) at different times. The current-state Jupiter is given by the light-blue curve at the age of the Solar System. The evolution of the heavy-element distribution (as a function of radius and mass) is available as \underline {online movies}.}\label{Z_Z}
   \end{figure}

Fig.~\ref{Z_Z} shows the primordial heavy-element distribution as a function of normalized mass/radius and its time evolution. The current-state Jupiter is shown in the light-blue line. 
As can be seen from the figure, at early stages, the vigorous convection "erodes" the outer part of the gradient and mixes it with the planetary upper envelope, which enriches the atmosphere with heavy elements.
Most of the mixing occurs during the first few 10$^7$ years when the contraction is efficient and the planet cools rapidly \citep{guilsteven04}.  
As the outer region of the planet cools more efficiently (than the inner region) by convection, the temperature gradient between the inner and the outer parts increases and small composition stairs are formed. However, further mixing of heavy elements is negligible. 
We find that for Jupiter today $\sim$\,60\% of the outer envelope becomes homogeneously mixed, with a heavy-element mass content of $\sim$\,16\me, i.e., metallicity of Z=0.08 in the outer envelope. 
Since all the heavy elements are represented by water, it is expected that the actual Z value will be somewhat lower consistent with the Galileo probe measurements \citep{wong2004}. 
The innermost regions are stable against convection and the steep composition gradient persists. 

  \begin{figure}
\centerline{\includegraphics[angle=0, width=9.6cm]{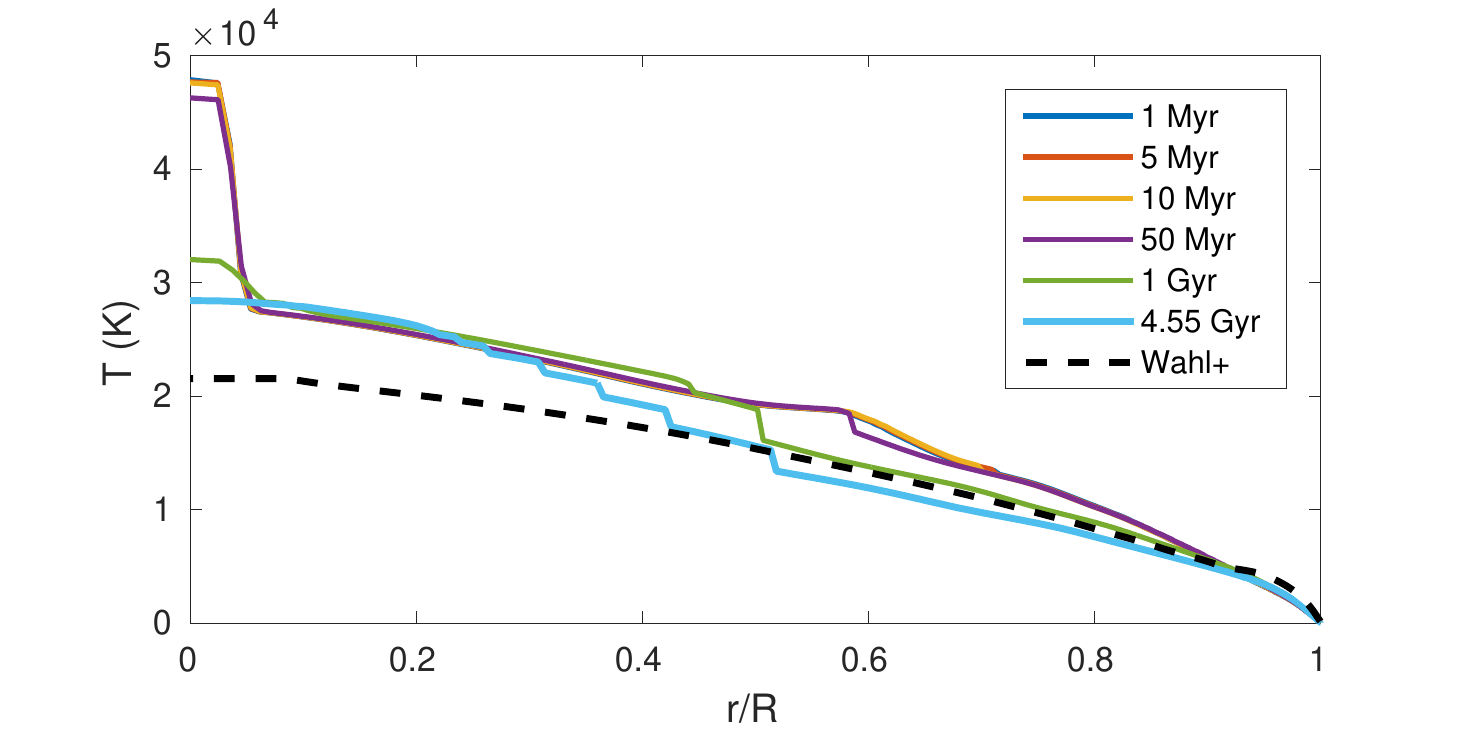}}
\vspace{-0.50ex}
\centerline{\includegraphics[angle=0, width=9.6cm]{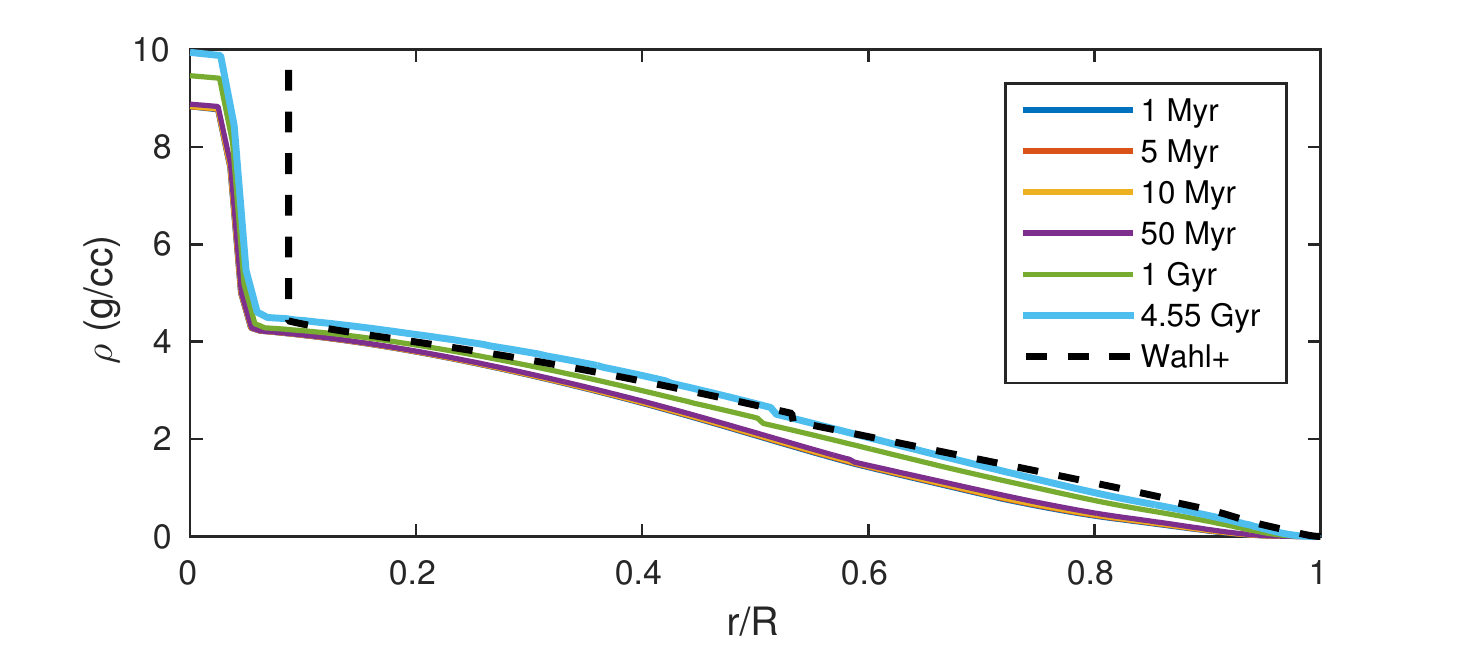}}
\vspace{-0.5ex}
\caption{Temperature (top) and density (bottom) vs. normalized radius at different ages. The different colors are for different times during Jupiter's evolution, as appears in the legend. The black dashed lines represents the profiles of Jupiter today assuming an adiabatic interior \citep{wahl17}. The evolution of the temperature (as a function of radius and mass) is available as \underline {online movies}}\label{T_T}
   \end{figure}
  
\begin{figure}
\centerline{\includegraphics[angle=0, width=9.5cm]{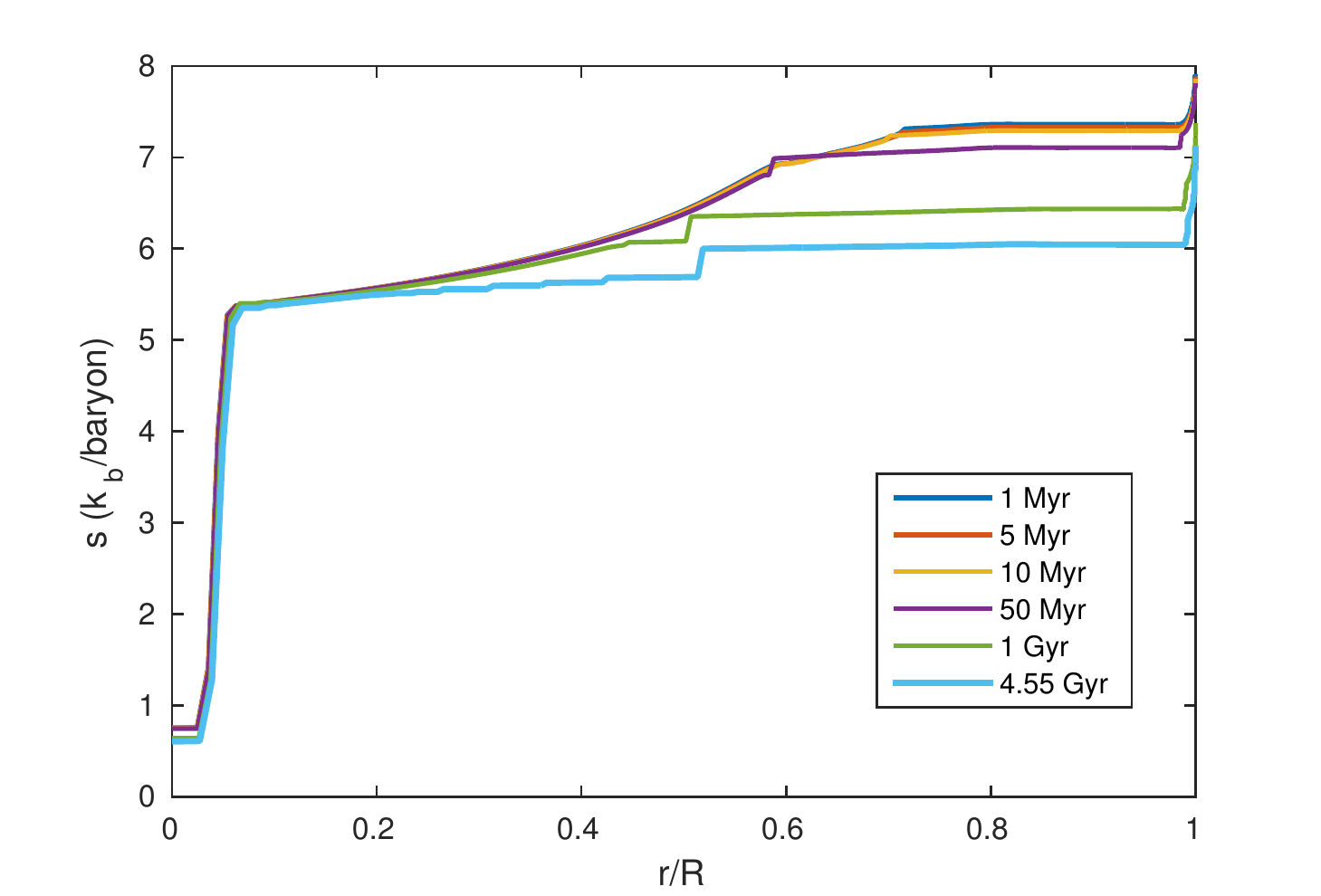}}
\vspace{-0.5ex}
\centerline{\includegraphics[angle=0, width=9.5cm]{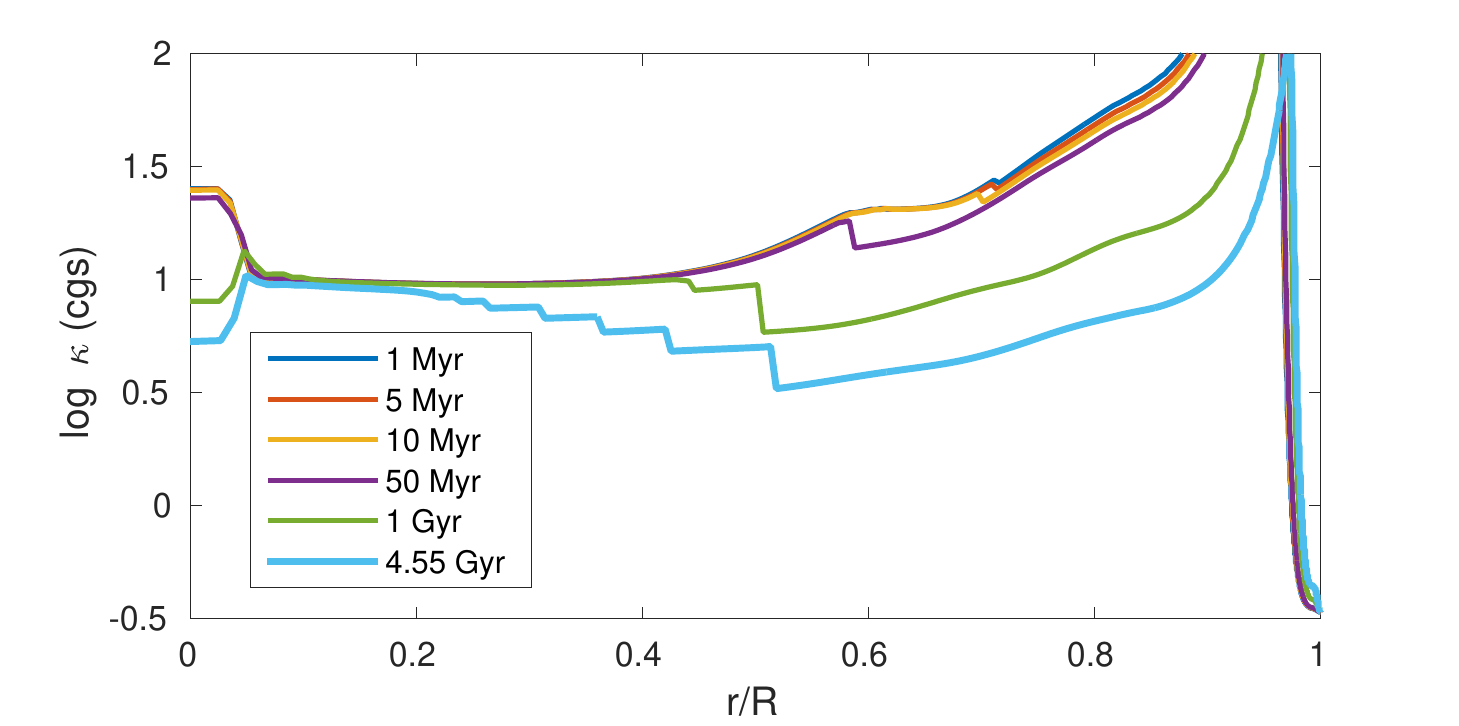}}
\vspace{-1.0ex}
\caption{Entropy (top) and opacity (bottom) vs. normalized radius at different times. The different colors are for ages as appears in the legend. 
The evolution of the entropy and opacity (as a function of radius and mass) are available as \underline {online movies}}
\label{rho_s}
   \end{figure}

Fig.~\ref{T_T} shows the temperature (top) and density (bottom) profiles at different times. For comparison, we also show the temperature and density profiles derived assuming an adiabatic interior by \cite{wahl17}. 
Since the outer region of the planet is mixed, the temperature profile overlaps with the adiabatic one. The inner region on the other hand, is significantly hotter, and the central temperature can be as high as 30,000 K, at present-day.  
In terms of density, as expected, the density increases as the planet evolves and contracts. The agreement for the current-state Jupiter is good but not excellent, this is likely to be linked to the SCVH EOS we are using in this model. Nevertheless, the fact that the density profile is very similar suggests that also a non-adiabatic interior of Jupiter can be consistent with observational constraints. 
Fig.~\ref{rho_s} shows the evolution of the entropy (top) and opacity (bottom). 
As the planet evolves the entropy decreases. The entropy decrease in the outer part is for two reasons: the cooling of the envelope and its heavy-element enrichment by convective-mixing. As convection progresses inward the adiabatic region of the envelope expands. The entropy in the innermost regions, on the other hand, is changing with depth due to the composition gradients (increasing in Z) and is  almost unchanged during the long-term evolution.

\subsection{The Appearance of Stairs}  
The appearance of stairs is an important aspect of this work and may be understood as follows: 
the outer regions tend to cool faster because in the primordial model this region is nearly metal-free and convectively unstable. The entropy decrease in the outer part is due to the cooling of the envelope and its heavy-element enrichment by convective-mixing. As convection progresses inward the adiabatic region of the envelope expands. The entropy in the innermost regions, on the other hand, is changing with depth due to the composition gradients (increasing in Z) and is  stable against convection.
Because of the cooling of the outer part, the transition region between the outer convective envelope and the stable inner region with composition gradients is characterised by a large jump in composition {\em and} in temperature, and the boundary between these two regions is progressively destabilised and moves inward.
\par

This smooth evolution however eventually changes and stairs appear due to a combination of three effects. 
First, the outer temperatures decrease continuously which reduces the cooling rate of the outer layer. 
Second, in the inner part of the planet, the increase in temperature and pressure implies that more electrons are available and heat is mostly transported by conduction. 
Third, the inner layers cool by conduction at an increasing rate due to the larger temperature jump at the transition and the lower conductive opacities (increased conduction) at deeper levels. 
When the rate at which the transition moves inward becomes smaller than the rate at which the inner region looses entropy, an inner convective zone appears. 
The temperature jump at the transition is then maintained at a level which makes it stable against convection and an entropy jump appears. The transition then stops progressing inward in mass. 
At deeper levels however the process continues: a convective zone appears and grows inward until the entropy loss becomes smaller than that of the even inner layers. Then a second staircase occurs. 
This process continues until the rate at which the conductivity increases becomes too small and/or the specific entropy decrease becomes too large.
\par 

Since the formation of stairs is linked to the conductivity (and $F_{cond} \propto dT/dr$), the exact location and size of the stairs depend on the number of grid points of the model. 
Nevertheless, we argue that the stairs are physical although their exact number (and size) is not well-determined. 
The large outer stairs have sizes that are larger than the pressure scale height $H_p$ while the smaller ones have sizes comparable to $H_p$ and in principle could be mixed by overshooting, where mixing extends beyond the convective region \citep[e.g.,][]{herwig97}. 
The region that is dominated by stairs is the place where the planet is expected to develop double-diffusive convection (DDC) as discussed below.

\subsection{Double-diffusion convection}

DDC can occur in regions that are found to be stable against convection according to the \ldx criterion, but unstable according to \swr criterion \citep{rosenblum11,wood13}. 
In these regions, the heat transport rate in our model is lower than in the case of DDC, since we treat these regions as
being radiative/conductive. 
Including DDC in evolution models requires knowledge of thermodynamical properties which are not well known such as the Prandtl number and the diffusivity ratio, as well as the assumed number of convective-diffusive layers \citep{lecontechab12}. 
Nevertheless, we can estimate the regions that are expected to develop DDC from the density ratio $R_{\rho}\equiv (\nabla_T-\nabla_{ad})/\nabla_{\mu}$ which is the ratio between the temperature gradient and the composition gradient \citep{mirouh12}.

\begin{figure}
\centerline{\includegraphics[angle=0, width=9.2cm]{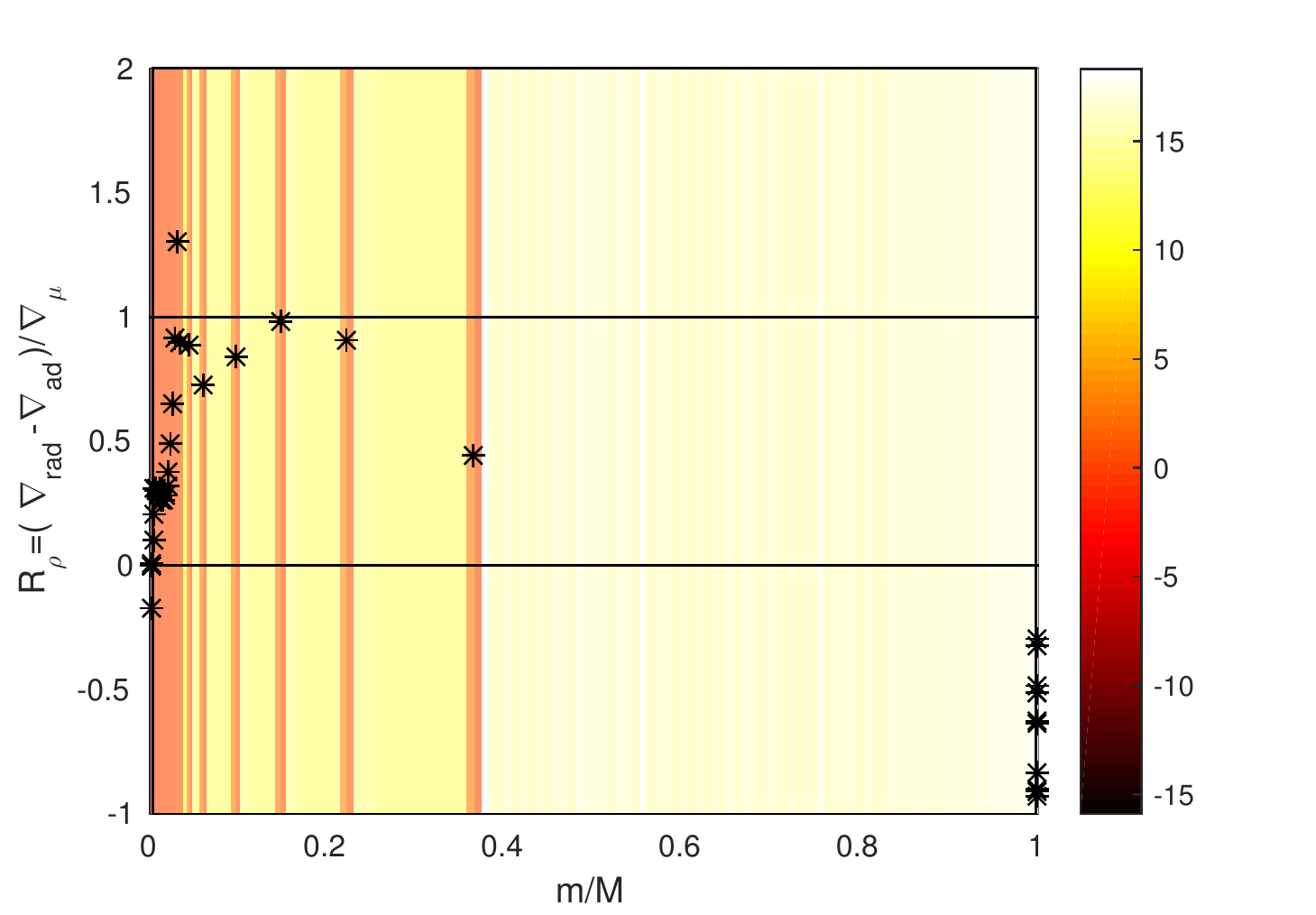}}
\vspace{-0.5ex}
\centerline{\includegraphics[angle=0, width=9.2cm]{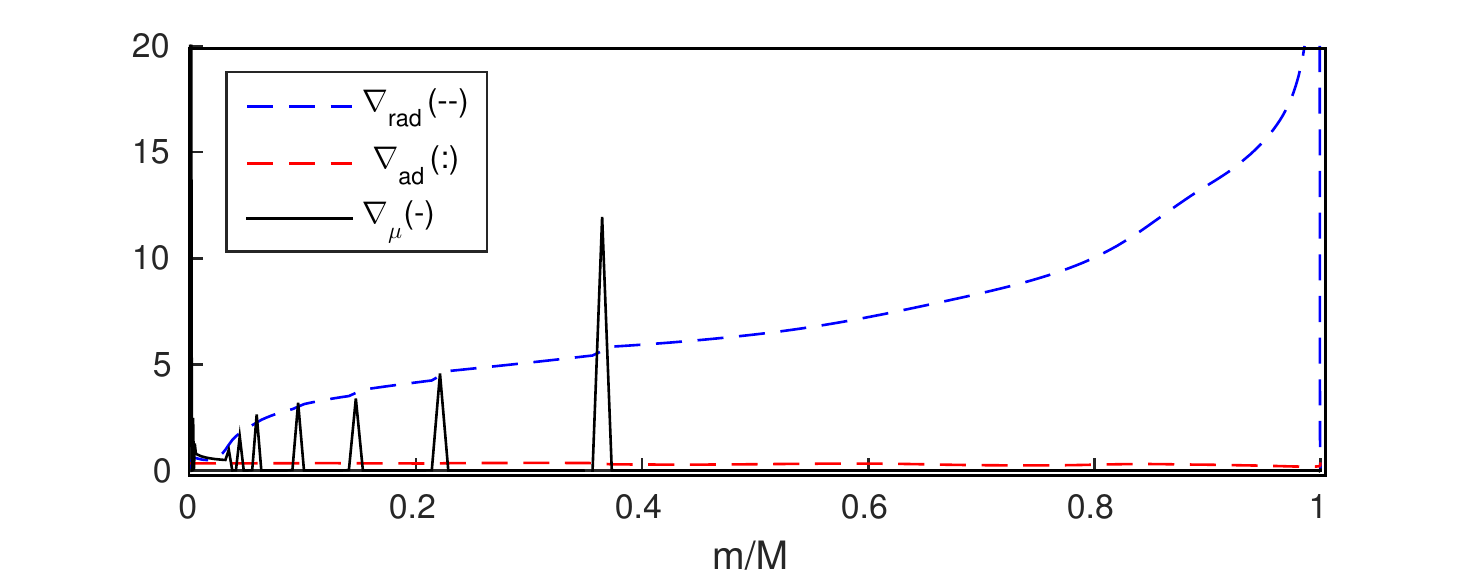}}
\vspace{-1.0ex}
\caption{{\bf Top}: the different heat transport regimes in the current-state Jupiter given by the density ratio $R_{\rho}$. Jupiter has three different regimes for the heat transport:  stable ($R_{\rho}<0$), DDC ($0<R_{\rho}<1$), and convective ($R_{\rho}>1$). Since the values of $R_{\rho}$ changes on a very wide range, the y-axis zoom-in the critical range, and the background colors represent the $\pm log(|R_{\rho}|)$.
{\bf Bottom}: The adiabatic, radiative and mean molecular weight gradients in the current-state model.}
\label{Rrho_nab}
   \end{figure}

Fig.~\ref{Rrho_nab} shows $R_{\rho}$ (upper panel) and the different temperature gradients (lower panel) for the current-state Jupiter. 
During the early evolution, we find that the innermost 30\% of the planet can develop layer-convection (not shown), while as the planet contracts and cools down, the DDC region shrinks and includes only 10\% of Jupiter's mass. 
We therefore conclude, that Jupiter's interior could have layered-convection, but the region is limited. However, in order to put robust limits on DDC we need to calculate the evolution self-consistently and explore the possible parameter space and their effect on the mixing efficiency.
Either way, DDC tends to soften the temperature gradient due to more efficient heat transport than the diffusive transport assumed here.

 \subsection{Sensitivity to the assumed heavy-element material, atmospheric opacity and H-He EOS}
Since rock has higher bulk density than ice, mixing upward of rock requires more energy. Therefore, hotter interiors are required to reach the same level of convective mixing \citep[see][for more details]{vazan15}. 
On the other hand, a smaller mass fraction of rock (compared to ice) is needed to produce the same envelope density profile. 
The outermost radiative zone, although negligible in terms of mass also affects the convective-mixing.
Therefore, we tested several radiative opacity calculations: we
use different factors of \cite{valencia13} as well as various factors of the \cite{pollack85}, and \cite{sharp07} opacities. 
We find that the main difference between the different opacity models is the atmospheric temperature distribution. We also find that  different radiative opacity values require different primordial temperatures to fit the current Jupiter parameters.
Since the opacity of Jupiter radiative zone is unknown, there is a range of possible primordial temperatures rather than one possibility. 
The EOS for H-He we use seems to be incompatible with more recent EOS calculations \citep[e.g.,][]{miguel16}. This can affect the inferred $M_Z$ but not the qualitative results of mixing. Future studies should perform a systematic analysis of the effects of the different assumptions on evolution and current-structure of Jupiter.

\section{Discussion and conclusions}
We presented the evolution of Jupiter assuming that its primordial structure consisted of composition gradients as suggested by planet formation models \citep{stevenson82a,lozovs17,helledsteven17}. 
We estimate $M_Z$ in Jupiter for a non-adiabatic structure with composition gradients to be $\sim$\,40\me. This value is consistent with the upper bounds derived by adiabatic models \citep{miguel16,wahl17}. 
We conclude that the maximal $M_Z$ in Jupiter is limited by convective-mixing.
Generally, in order to retain the measured radius and density profiles, higher $M_Z$ requires increase of internal temperatures.
However, in some point the temperature is high enough to initiate convection (and convective-mixing) which flattens the composition gradient, enhances the envelope cooling and enriches the envelope with heavy-elements, results in a decrease in the planetary radius. 
We therefore suggest that structure models must be consistent with the long-term evolution of the planet, since the two are linked. 
We also suggest that the enrichment of Jupiter's atmosphere with heavies can be explained by the existence of primordial composition gradients that slowly mixes during the planetary contraction. 
Follow-up work can include further complexities such as the sensitivity to the assumed H-He EOSs, various heavy elements, and different atmospheric opacities.

Here we assumed that composition gradients are primordial. Later-stage composition gradients such as helium rain, are more likely to survive because of the lower internal temperatures and the lower efficiency of convective mixing. 
In any case, helium rain in Jupiter is expected to occur in the region of our convective envelop. Therefore, Jupiter is expected to be separated to at least three different layers, in agreement with recent structure models \citep{miguel16,wahl17}. 

Finally, our study provides an independent method to estimate the heavy-element mass in a non-adiabatic Jupiter, and we suggest that such a non-standard configuration for Jupiter is not only consistent with observations, but is also predicted by formation models. 
Thus, structure and evolution models that are consistent with formation models are important for our understanding of the outer planets in our own planetary system and for the characterization of giant exoplanets.

\begin{acknowledgements}
 We thank the referee for valuable comments. A.V.\ acknowledges support by the Amsterdam Academic Alliance (AAA) Fellowship.  R.H.~
acknowledges support from the Swiss National Science Foundation (SNSF) Grant No. 200021\_169054
\end{acknowledgements}

\bibliographystyle{aa} 
\bibliography{allona.bib} 

\section*{A1. The planetary long-term evolution}
 
 The planetary evolution is modelled using a planetary evolution code that solves the structure and evolution equations accounting for convective mixing \citep[see][for details]{vazan15}.  
The heat transport mechanism is determined according to the \ldx convection criterion \citep{ledoux47} with material transport being computed as a convective-diffusive process in the convective regions which is given by:
$\nabla_R - \nabla_A - \nabla_X >0$,
where $\nabla_R$ and $\nabla_A$  are the radiative and adiabatic temperature gradients, respectively; and 
$\nabla_X=\sum_j\,[\partial \ln T(\rho,p,X)/{\partial X_j}]\,[{dX_j}/{d\ln p}]$
is the composition contribution to the temperature gradient.
For uniform composition, convection occurs when $\nabla_R>\nabla_A$, which is the standard \swr criterion \citep{schwarz06}. 
\par

 \begin{figure}
\centerline{\includegraphics[angle=0, width=9.2cm]{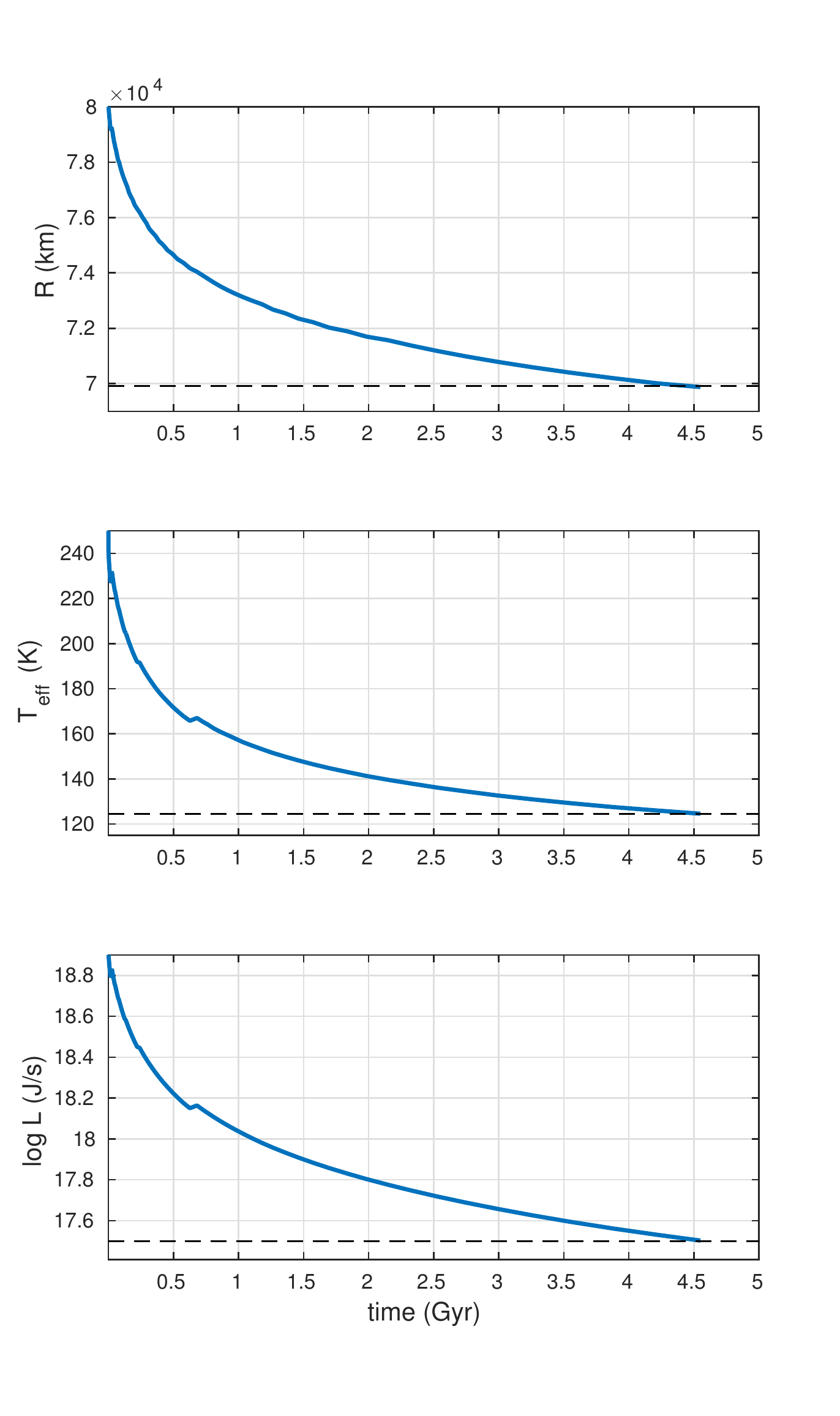}}
\caption{Radius (top), effective temperature (middle) and luminosity (bottom) evolution for our Jupiter model. The dashed curves correspond to the values measured at Jupiter today.}
\label{RTL_ev}
   \end{figure}

Figure~\ref{RTL_ev} shows the evolution of the radius, effective temperature, and luminosity of our Jupiter model. 
The inferred values at present day are consistent with the measured values (shown by the dashed curves). 
During the early evolution there are several small jumps in the evolution parameters, as a result of the efficient convetive-mixing progress inward. 
The jumps occur when vast convection changes the planetary structure, as discussed in the main text.

\end{document}